\documentclass[12pt]{article}

\usepackage{newtxtext,newtxmath}

\usepackage{graphicx}

\usepackage[letterpaper,margin=1in]{geometry}

\renewenvironment{abstract}
	{\quotation}
	{\endquotation}

\date{}

\makeatletter
\renewcommand{\fnum@figure}{\textbf{Figure \thefigure}}
\renewcommand{\fnum@table}{\textbf{Table \thetable}}
\makeatother

\usepackage{scicite}

\usepackage{url}

\usepackage{setspace}

\newcommand{\kms}{${\rm km\ s^{-1}}$}
\newcommand{\hi}{H{\sc i}}
\newcommand{\arcsec}{$^{\prime\prime}$}

\usepackage{xcolor}
\definecolor{fmcolor}{RGB}{250,50,50} 
\newcommand{\editfm}[1]{{#1}}

\usepackage{caption}
\def\scititle{
Supernova origin of galactic turbulence revealed by superbubbles
}
\title{\bfseries \boldmath \scititle}

\author{%
Fanyi Meng$^{1, 2\dagger}$, Chao-Wei Tsai$^{3, 4, 2\dagger}$, Jingwen Wu$^{2, 3\dagger}$,\and
Sihan Jiao$^{3, 5}$, Mordecai-Mark Mac Low$^{6}$, Zhi-Yu Zhang$^{7, 8}$,\and
Am\'{e}lie Saintonge$^{9}$, Hui Li$^{10}$, Zongnan Li$^{11}$, Jie Wang$^{3}$, Lile Wang$^{12, 13}$,\and
Haitao Xu$^{14}$, Yanbin Yang$^{15}$, Kai Zhang$^{3}$, Rouyu Li$^{2, 3}$, Di Li$^{1, 16\ast}$\and
\parbox[t]{0.96\textwidth}{\raggedright\footnotesize
$^{1}$New Cornerstone Science Laboratory, Department of Astronomy, Tsinghua University, Beijing 100084, China;
$^{2}$School of Astronomy and Space Science, University of Chinese Academy of Sciences, Beijing 100049, China;
$^{3}$National Astronomical Observatories, Chinese Academy of Sciences, 20A Datun Road, Beijing 100101, China;
$^{4}$Institute for Frontiers in Astronomy and Astrophysics, Beijing Normal University, Beijing 102206, China;
$^{5}$Max Planck Institute for Astronomy, K\"{o}nigstuhl 17, D-69117 Heidelberg, Germany;
$^{6}$Dept.\ of Astrophysics, American Museum of Natural History, 200 Central Park West, New York, NY 10024, USA;
$^{7}$School of Astronomy and Space Science, Nanjing University, Nanjing 210093, China;
$^{8}$Key Laboratory of Modern Astronomy and Astrophysics (Nanjing University), Ministry of Education, Nanjing 210093, China;
$^{9}$Max Planck Institute for Radio Astronomy (MPIfR), Auf dem H{\"u}gel 69, 53118 Bonn, Germany;
$^{10}$Department of Astronomy, Tsinghua University, Beijing 100084, China;
$^{11}$Korea Astronomy and Space Science Institute, 776 Daedeok-daero, Yuseong-gu, Daejeon 34055, Republic of Korea;
$^{12}$The Kavli Institute for Astronomy and Astrophysics, Peking University, Beijing 100871, China;
$^{13}$Department of Astronomy, School of Physics, Peking University, Beijing 100871, China;
$^{14}$Center for Combustion Energy and School of Aerospace Engineering, Tsinghua University, Beijing, 100084, China;
$^{15}$LIRA, Observatoire de Paris, Universite PSL, CNRS, Place Jules Janssen 92195, Meudon, France;
$^{16}$State Key Laboratory of Radio Astronomy and Technology, National Astronomical Observatories, Chinese Academy of Sciences, Beijing 100101, China.\\
$^\dagger$ These authors contributed equally to this work.\\
$^\ast$ Corresponding author. Email: dili@tsinghua.edu.cn}%
}

\usepackage[hidelinks,hypertexnames=false]{hyperref}
\hypersetup{pdftitle={Supernova origin of galactic turbulence revealed by superbubbles},
pdfsubject={Initial-submission manuscript; revised Version of Record in Nature Astronomy},
pdfauthor={Fanyi Meng and coauthors}}
\begin{document} 

\maketitle
\begingroup
\small\setlength{\parskip}{4pt}\setlength{\parindent}{0pt}
\textbf{Version notice: initial-submission manuscript.}
This preprint has not undergone peer review (when applicable) or any post-submission improvements or corrections.
The Version of Record of this article is published in \textit{Nature Astronomy}, and is available online at \url{https://doi.org/10.1038/s41550-026-02981-9}.

Revisions made in response to peer review are not incorporated in this manuscript.
The revised Version of Record was published on 17 September 2026.
Please consult and cite that version for the final results, methods and conclusions.
\endgroup

\clearpage

\begin{abstract} \bfseries \boldmath

Supernovae (SNe) are among the leading candidates for powering galactic-scale turbulence.
SNe drive expanding shells of neutral atomic hydrogen (HI) known as superbubbles.
Due to the lack of a sensitive, dynamically complete, galaxy-wide census, superbubbles have not been used to quantify the galactic-scale turbulent energy budget. 
Here we present a combined Five-hundred-meter Aperture Spherical radio Telescope (FAST) and Jansky Very Large Array HI survey of the Andromeda galaxy (M31), the nearest giant spiral, with superior sensitivity and dynamical coverage. 
We identify 118 superbubbles across the entire disk of M31 with dynamical ages up to 40 Myr, consistent with the expected duration of SN activity in a star cluster and extending the age coverage well beyond previous surveys.
Inferred from these superbubbles, the kinetic energy injection rates ($10^{49}$–$10^{51.5}$ erg kpc$^{-3}$ Myr$^{-1}$) from SNe closely match the turbulence dissipation rates derived independently from the same data, in both magnitude and spatial distribution. 
These results demonstrate that clustered SN feedback is sufficient to sustain galactic-scale turbulence, which shapes disk structure and influences galaxy evolution.
\end{abstract}

\noindent

Neutral gas in galaxies exhibits characteristic velocity dispersions that exceed the thermal sound speed and are widely attributed to turbulence \cite{Elmegreen:2004uc,Mac-Low:1999ue,Utomo:2019uk}.
Acting across a wide range of spatial scales, this turbulence shapes galactic structure, retards the gravitational collapse of atomic gas into molecular clouds, and regulates the rate and efficiency of star formation \cite{Mac-Low:2004te,Padoan:2011aa,Padoan:2020aa}.
On galactic scales, it is most effectively traced by the neutral atomic hydrogen (\hi), where velocity dispersions of 10--20\,km\,s$^{-1}$ are commonly observed \cite{Petric:2007aa,Tamburro:2009uz}.
Galactic-scale turbulence dissipates rapidly, with a characteristic decay timescale of $\sim$10\,Myr and a volumetric energy loss rate of $\dot{e}_{\rm turb} \sim 10^{-27}$\,erg\,cm$^{-3}$\,s$^{-1}$ \cite{Stone:1998aa,Mac-Low:1999ue}.
The dominant energy source to replenish such energy loss has remained uncertain for decades \cite{Sellwood:1999aa,Wada:2002aa,Krumholz:2016wn}.
Proposed mechanisms include stellar feedback, magnetorotational instability (MRI), and large-scale mass transport within galactic disks \cite{Krumholz:2016wn,Beattie:2025aa}.
Among these, SNe have long been regarded as a leading candidate \cite{Tamburro:2009uz,Bacchini:2020aa,Mac-Low:2004te}.
Indirect estimates based on star formation rates suggest that the energy input from SNe exceeds the turbulent dissipation rate by one to two orders of magnitude \cite{Mac-Low:1999ue,Mac-Low:2004te,Joung:2006aa}.
These estimates rely on assumptions about the stellar initial mass function, SN timescales, and the efficiency of kinetic energy coupling to the interstellar medium.
Whether SNe are indeed the primary energy source of galactic turbulence, and how they power and sustain it, has remained uncertain owing to the lack of direct observational evidence.

Here we show that superbubbles can be used to directly trace how SNe inject energy into galactic turbulence.  
Superbubbles are large gas shells and cavities, typically $\gtrsim100$\,pc in size.  
They are most commonly identified in \hi\ emission and are understood to be driven by clustered, sequential SNe \cite{Heiles:1979we,Brinks:1986vn,Bagetakos:2011ub,Suad:2019aa,Li:2024aa}.  
Because their radii, expansion velocities, and characteristic expansion timescales can be measured directly, superbubbles provide a way to quantify the rate at which SNe deposit kinetic energy into the interstellar medium.   
Molecular or dust tracers can identify some superbubbles \cite{Watkins:2023wj}, 
while \hi\ provides complementary sensitivity to older, more diffuse structures \cite{Watkins:2023tk} 
and carries most of the kinetic energy of the neutral gas (see below).
High-resolution \hi\ observations are therefore essential for measuring how efficiently SNe inject energy into the surrounding gas.  

\begin{figure*}[!ht]
\centering{
\textsf{
\includegraphics[width=0.85\textwidth]{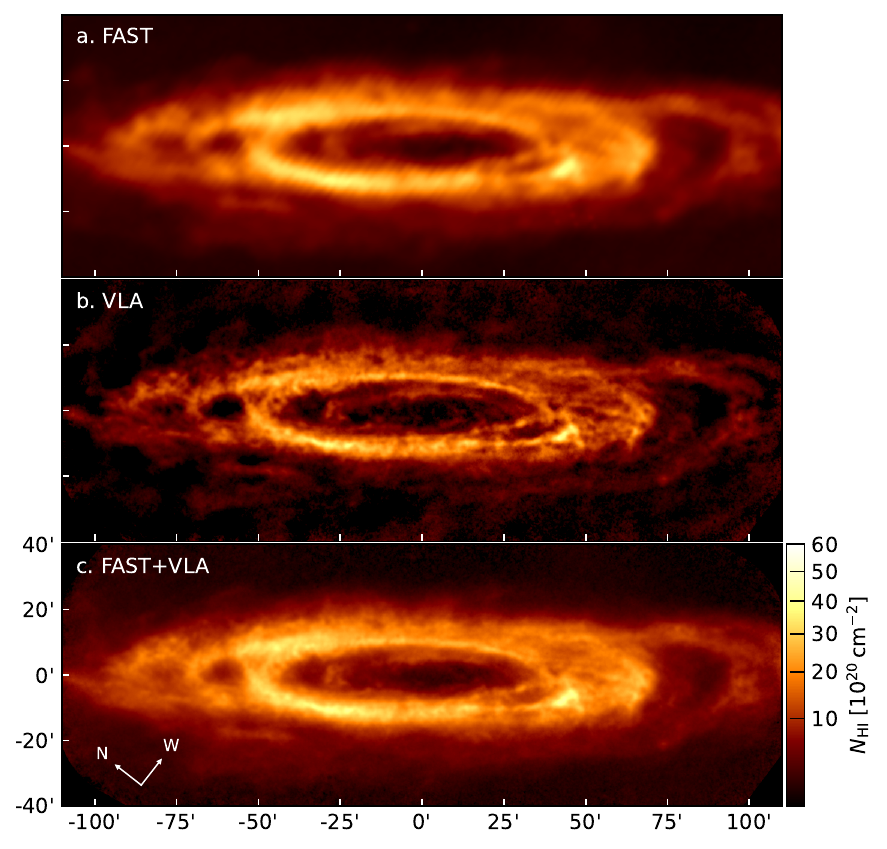}
\caption{\textbf{\hi\ column density maps of M31 from FAST, JVLA, and their combination.}
Each panel displays the 21-cm moment-0 map converted to atomic hydrogen column density, centered on the M31 disk. The white arrows in the lower-left corner of panel (c) indicate the north (N) and west (W) directions on the sky. The beam size in panel (a) is 3$^{\prime}$ ($\sim 650\,{\rm pc}$) and in panels (b) and (c) is 1$^{\prime}$ ($\sim 216\,{\rm pc}$). \label{f.three}}
}}
\end{figure*}

\begin{figure*}[!ht]
\centering{
\textsf{
\includegraphics[width=0.59\textwidth]{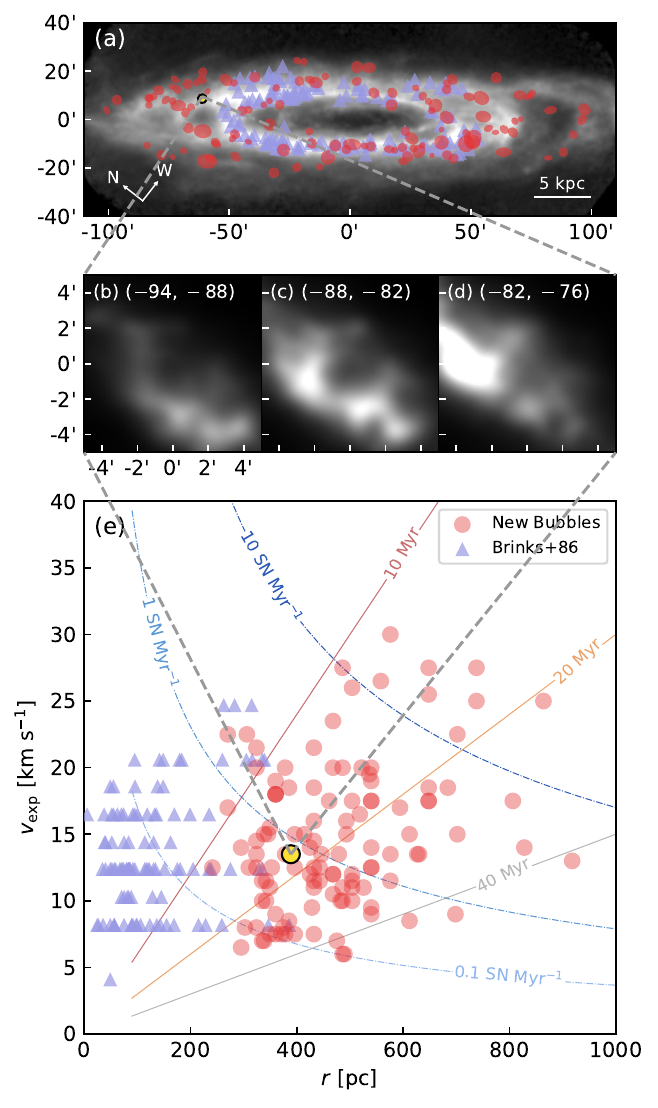}
\caption{
\textbf{Spatial distribution and properties of the \hi\ superbubbles in M31.}
(a) \hi\ column density map of M31 with ellipses marking the identified superbubbles in our sample (red) and those (blue) reported by Brinks+86 \cite{Brinks:1986vn}.  
(b–d) \hi\ column density maps for a representative superbubble, integrated over three velocity ranges (labeled in each panel in ${\rm km\,s^{-1}}$), illustrating the velocity span of the structure.  
The location of this bubble is indicated by a yellow circle in panels (a) and (e).  
(e) Distribution of our superbubbles (red) in the $r$–$v_{\rm exp}$ plane.  Bubbles reported by Brinks+86 are in blue.
The dashed curves show the expected $r$–$v_{\rm exp}$ relations for constant SN rates computed assuming $n_{\rm HI}=0.2\ {\rm cm^{-3}}$ \cite{Ehlerova:2013uh}, and the solid lines show iso-timescale tracks for expansion times of 10, 20, and 40~Myr based on the Weaver model \cite{Weaver:1977wh}.  
\label{f.momentzero}
}
}
}
\end{figure*}

We obtained the most sensitive \hi\ observations of the Andromeda galaxy (M31) to date by combining new data  from the Five-hundred-meter Aperture Spherical radio Telescope (FAST), which is described by Ref.\ \cite{Li:2018aa}, with archival Jansky Very Large Array (JVLA) observations \cite{Koch:2021vf}.  
M31, at a distance of 0.75\,Mpc \cite{Vilardell:2010vm}, is close enough to resolve individual superbubbles while avoiding the line-of-sight confusion that affects studies within the Milky Way.  
The resulting FAST--JVLA cube reaches a root-mean-square column density sensitivity of $\sim3\times10^{17}\,{\rm cm^{-2}}$ per $1.68\,{\rm km\,s^{-1}}$ channel at a spatial resolution of $\sim216\,{\rm pc}$.  
This high sensitivity allows us to maintain fine enough spectral resolution to detect slowly expanding, evolved superbubbles with expansion timescales up to $\sim40$\,Myr, comparable to the full duration of SN activity in a massive stellar cluster \cite{Orr:2022aa}.  
Earlier \hi\ surveys from four decades ago, constrained by sensitivity and/or velocity resolution \cite{Brinks:1986vn}, were primarily sensitive to younger bubbles concentrated in the `10\,kpc ring' \cite{Block:2006uz,Chemin:2009vo}, as shown in Fig.~\ref{f.momentzero}.  
The new observations extend sensitivity to late evolutionary phases of superbubbles, enabling a time-averaged estimate of the kinetic energy injection rate, $\dot{e}_{\rm bubble}$, that can be directly compared with the local turbulence dissipation rate, $\dot{e}_{\rm turb}$.
Incorporating FAST data is also crucial for accurately measuring the \hi\ line width and thus reliably estimating $\dot{e}_{\rm turb}$, as FAST recovers spatially extended emission with intrinsically broader profiles \cite{Yue:2021aa,Plunkett:2023aa}, as detailed in Materials and Methods.
Together, these datasets enable a joint analysis of superbubble energetics and interstellar turbulence in M31. The images of the FAST, JVLA, and combined data are shown in Fig.~\ref{f.three}

Across the M31 disk, we identified 118 \hi\ superbubbles.  
For each bubble, the radius $r$ and expansion velocity $v_{\rm exp}$ were measured directly from the data cube.  
The expansion timescale was then estimated as $t_{\rm exp}=0.6\,r/v_{\rm exp}$ \cite{Weaver:1977wh,Silich:1996vt}.  
The ambient \hi\ volume density $n_{\rm HI}$ was derived from the local \hi\ column density together with the scale height of M31's \hi\ disk \cite{Braun:1991ty}.  
Figure~\ref{f.momentzero} shows the bubble locations and their measured properties.  
Details of the identification procedure and parameter estimation are provided in Materials and Methods.  

These superbubbles are produced by SN activity.  
Galactic shear can generate cavities in differentially rotating disks \cite{Wada:2000aa,Wada:2002aa}, but all measured expansion velocities exceed those expected from shear, ruling out a shear origin (see Fig.~\ref{f.nsf} and Materials and Methods).  
Other feedback mechanisms cannot account for the large sizes of the bubbles we detect ($r > 200\,{\rm pc}$), whereas simulations show that SNe can produce bubbles of this scale \cite{Li:2024aa}.  
The total energies required to drive the superbubbles, estimated using the Weaver model together with more recent modeling \cite{Weaver:1977wh,El-Badry:2019tf}, all exceed the canonical $10^{51}\,\mathrm{erg}$ released by a single SN \cite{McKee:1977wl,Burrows:2013aa}, indicating that most bubbles arise from multiple explosions (see Fig.~\ref{f.rvage}).  
The momentum per SN, inferred from the bubble momenta and the SN counts derived above, is approximately $10^{5.5\text{--}6.5}\,M_{\odot}\,{\rm km\,s^{-1}}$ (see Fig.~\ref{f.momentum} and Materials and Methods), consistent with theoretical predictions \cite{Kim:2015aa,Kim:2017ab,Gentry:2017aa}.  
The bubbles also have expansion timescales of $10$--$40\,{\rm Myr}$, matching the duration of SN activity in massive stellar clusters \cite{Leitherer:1999aa,Orr:2022aa}.

\begin{figure}[!ht]
\centering{
\textsf{
\includegraphics[width=0.465\textwidth]{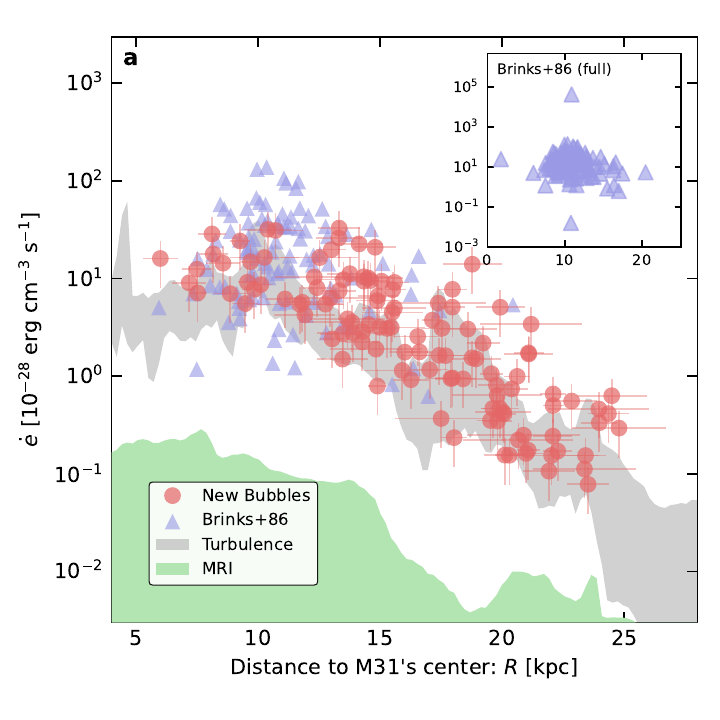}
\includegraphics[width=0.465\textwidth]{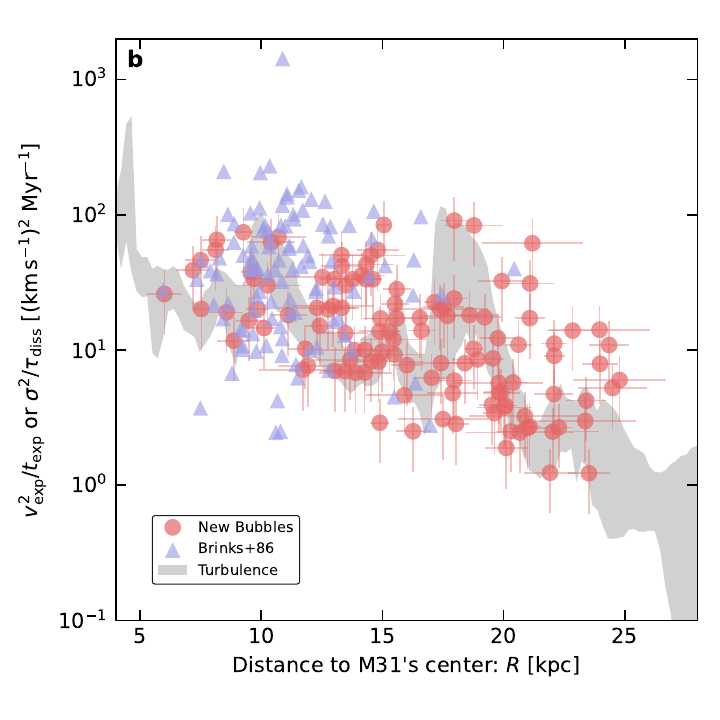}\\
\includegraphics[width=0.465\textwidth]{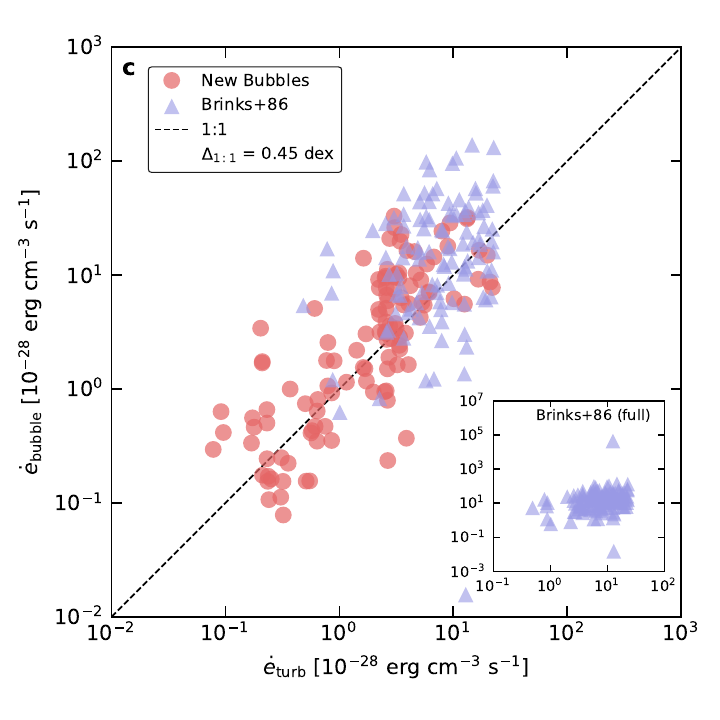}
\includegraphics[width=0.465\textwidth]{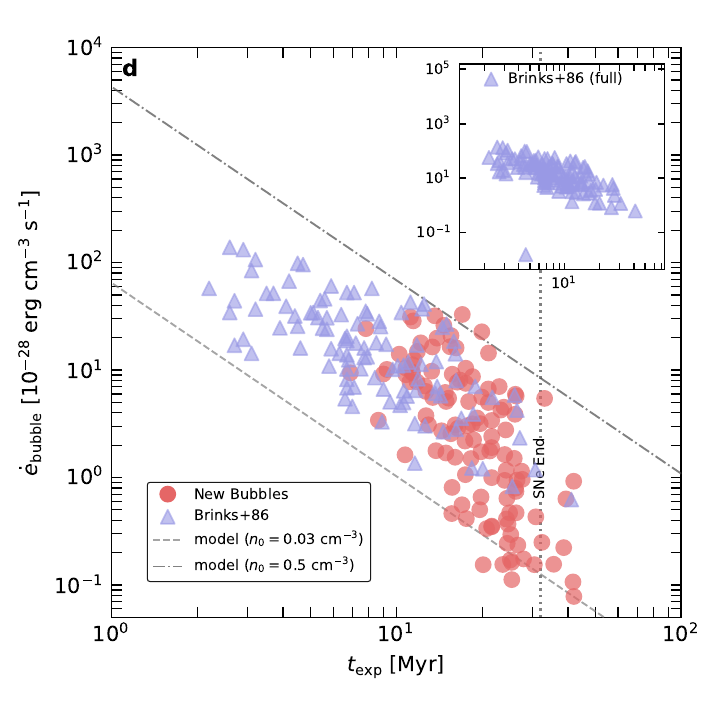}
\caption{
\textbf{Superbubble energy injection compared with turbulent dissipation.}
(a) Volumetric energy injection rates $\dot{e}_{\rm bubble}$ for individual bubbles (red) as a function of $R$, shown alongside the turbulent dissipation rate (gray band) and the MRI estimate (green shade); Brinks+86 shells \cite{Brinks:1986vn} are plotted in blue, and the inset shows their full range.
(b) Radial comparison of $v_{\rm exp}^2/t_{\rm exp}$ for bubbles with $\sigma^2/\tau_{\rm diss}$ for turbulence (gray band), using the same color scheme as panel a for both the new bubbles and the Brinks+86 shells \cite{Brinks:1986vn}.
(c) Comparison of $\dot{e}_{\rm bubble}$ and $\dot{e}_{\rm turb}$ at matched $R$ relative to a 1:1 line, with a median offset of 0.45\,dex; the inset shows the full Brinks+86 sample \cite{Brinks:1986vn}.
(d) $\dot{e}_{\rm bubble}$ versus expansion time, compared with Weaver-model (kinetic energy part) tracks for $n_0=0.03$ ($r_0=50\,{\rm pc}$ at $t_{\rm exp}=0$) and $0.5\,{\rm cm^{-3}}$ ($r_0=100\,{\rm pc}$ at $t_{\rm exp}=0$); the dotted line marks the end of clustered SNe \cite{Leitherer:1999aa,Orr:2022aa}, and Brinks+86 bubbles are overplotted \cite{Brinks:1986vn}.
\label{f.mainchart}
}
}}
\end{figure}

We quantified the time-averaged volumetric rate of kinetic energy injection by superbubbles using  
\begin{equation}\label{eq.dotebubble}
     \dot{e}_{\rm bubble} = \frac{E_{\rm k}}{V\,t_{\rm exp}} = \frac{1}{2}\frac{\mu_{\rm H} n_{\rm HI}\,v_{\rm exp}^2}{t_{\rm exp}},
\end{equation}
where $V = (4/3)\pi r^3$ is the bubble volume.  
The ambient density $n_{\rm HI}$ was measured in an annulus of thickness $r$ surrounding each bubble.  
We assumed isotropic expansion into a locally homogeneous \hi\ medium of density $n_{\rm HI}$ and adopted a mean mass per hydrogen atom, including helium, of $\mu_{\rm H} = 2.34\times10^{-24}\,{\rm g}$ \cite{Krumholz:2009aa}.  
This estimate depends solely on observed bubble properties and accounts only for their kinetic energy.  
As shown in Fig.~\ref{f.mainchart}, at galactocentric radii $R = 15$--$20\,{\rm kpc}$, $\dot{e}_{\rm bubble}$ declines from  
$\sim 3\times10^{-27}$ to $\sim 10^{-29}\,{\rm erg\,cm^{-3}\,s^{-1}}$  
(or $\sim 10^{51.5}$ to $\sim 10^{49}\,{\rm erg\,kpc^{-3}\,Myr^{-1}}$).  
Although bubbles in the inner and outer disk have similar radii, those in the outer disk expand more slowly, are generally older, and reside in regions where $n_{\rm HI}$ decreases with radius (Fig.~\ref{f.rvage}).  
These trends together explain the radial decline in $\dot{e}_{\rm bubble}$.  
\editfm{Potential interactions among superbubbles could in principle alter their evolution.  
In our sample, 42 of 118 bubbles show some projected overlap and the shared area represents just 4.6\,\%, implying that true three-dimensional encounters are even rarer.
Where interactions do occur, the resulting deceleration increases $t_{\rm exp}$ and therefore renders our estimate of $\dot{e}_{\rm bubble}$ conservative.  
A more precise assessment of the interaction probability would require detailed constraints on the local diffuse FUV heating, which regulates how freely bubbles expand \cite{Hill:2018wi}. } 

\editfm{We also computed $\dot{e}_{\rm bubble}$ for superbubbles from previous \hi\ observations of M31 \cite{Brinks:1986vn} (hereafter Brinks+86).  
Although that study focused on estimating the total energy required to form each shell, the reported radii, expansion velocities, and ambient densities equally allow the kinetic-energy component of $\dot{e}_{\rm bubble}$ to be derived in a manner consistent with our sample.  
In doing so, we recalculated their $n_{\rm HI}$ using the updated \hi\ scale height $h$ from \cite{Braun:1991ty}, and the resulting values are also shown in Fig.~\ref{f.mainchart}.  
Brinks+86 bubbles, concentrated in the `10 kpc ring' (see Fig.~\ref{f.momentzero}), generally show higher $\dot{e}_{\rm bubble}$ than our bubbles and than $\dot{e}_{\rm turb}$ at their corresponding $R$ (see Fig.~\ref{f.mainchart}).  
Such differences likely arise from the distinct evolutionary stages sampled by the two surveys.  
As shown in Fig.~\ref{f.mainchart}(d), the Brinks+86 bubbles are typically younger by $\sim 0.5$\,dex (10--20\,Myr), which, under the Weaver model \cite{Weaver:1977wh}, leads to values of $\dot{e}_{\rm bubble}$ that are higher by a factor of $\sim 3.5$.  
In contrast, most bubbles in our sample have $t_{\rm exp}$ values already approaching the limit of SNe activity in a stellar cluster \cite{Leitherer:1999aa}, indicating that our sample traces the almost complete mechanical output of clustered SNe.  }

We calculated the turbulence dissipation rate as  
\begin{equation}\label{eq.doteturb}
    \dot{e}_{\rm turb} = \frac{3}{2}\,\frac{\mu_{\rm H} n_{\rm HI}\sigma_{\rm HI}^2}{\tau_{\rm D}},
\end{equation}
where $\sigma_{\rm HI}$ is the non-thermal velocity dispersion and $\tau_{\rm D} = 2h/\sigma_{\rm HI}$ is the dissipation timescale, assuming that the largest eddies span the \hi\ disk scale height $h$ \cite{Tamburro:2009uz,Bacchini:2020aa}.  
The factor of $3/2$ reflects the assumption of isotropic turbulence.  
The non-thermal dispersion $\sigma_{\rm HI}$ was obtained from the second moment after subtracting in quadrature the thermal component, estimated by assuming that 60\,\% of the \hi\ resides in the warm neutral medium with a thermal width of $8\,{\rm km\,s^{-1}}$ \cite{Dickey:1993ue}.  
We included only pixels where projection broadening is smaller than the spectral resolution (see Materials and Methods and Fig.~\ref{f.deltav_model}).  
As shown in Fig.~\ref{f.mainchart}, $\dot{e}_{\rm turb}$ shows the same radial decline as $\dot{e}_{\rm bubble}$, decreasing from  
$\sim 3\times10^{-27}$ to $\sim 10^{-29}\,{\rm erg\,cm^{-3}\,s^{-1}}$  
(or $\sim 10^{51.5}$ to $\sim 10^{49}\,{\rm erg\,kpc^{-3}\,Myr^{-1}}$) over $R = 15$--$20\,{\rm kpc}$.  
This decline arises primarily because $n_{\rm HI}$ decreases with radius while $\sigma_{\rm HI}$ remains nearly constant (Fig.~\ref{f.m31_profile}). 

Overall, $\dot{e}_{\rm bubble}$ for our bubbles and $\dot{e}_{\rm turb}$ agree closely in both their absolute values and radial behavior (see Fig.~\ref{f.mainchart}a and Fig.~\ref{f.mainchart}c).  
\editfm{To ensure that this correspondence is not merely driven by the large dynamic range of $n_{\rm HI}$, we compare $v_{\rm exp}^2/t_{\rm exp}$ with $\sigma_{\rm HI}^2/\tau_{\rm D}$ in Fig.~\ref{f.mainchart}b, which removes the explicit density dependence.  
These ratios remain consistent in both magnitude and radial trend, demonstrating that the agreement between $\dot{e}_{\rm bubble}$ and $\dot{e}_{\rm turb}$ reflects a genuine physical connection.}  
Although both quantities are derived from the same \hi\ data set, they rely on different measurements and assumptions and are sensitive to distinct sources of uncertainty.  
Their close correspondence therefore indicates that SNe, acting through superbubbles, are sufficient to sustain the observed \hi\ turbulence.

We focus only on the \hi\ component of $\dot{e}_{\rm bubble}$ and $\dot{e}_{\rm turb}$.  
For molecular gas, superbubbles in our sample are older than $10\,{\rm Myr}$, a timescale over which molecules are destroyed \cite{Bagetakos:2011ub,Watkins:2023wj}.  
In addition, within the inner $12\,{\rm kpc}$ of M31's disk, the radial extent of CO detection \cite{Nieten:2006tr}, the molecular gas accounts for $\lesssim 10\%$ of the total neutral gas mass (see Fig.~\ref{f.moltohiratio}).  
This fraction also holds at the specific locations of our bubbles within this radius, indicating that our bubble sample does not preferentially trace regions with above-average molecular content, as detailed in the Materials and Methods.
Estimates of turbulence energy in nearby galaxies further show that the contribution of molecular turbulence to the kinetic energy budget is negligible compared to that of atomic gas within our uncertainty \cite{Hughes:2013ut,Tamburro:2009uz}.  
For ionized gas, most massive stars have already exploded as SNe by $10\,{\rm Myr}$ and can no longer sustain ionization \cite{Watkins:2023wj}, so the ionized gas associated with the bubbles can be neglected.  
To estimate the possible contribution of the warm ionized medium (WIM) to $\dot{e}_{\rm turb}$, we adopt a WIM-to-\hi\ volume density ratio of $\sim 0.1$ from Milky Way measurements \cite{Hill:2008aa,Haffner:2009aa}, assume a WIM scale height about three times larger than that of \hi, and use a velocity dispersion of $\sim 30\,{\rm km\,s^{-1}}$ measured in the central $<5\,{\rm kpc}$ of M31 \cite{Opitsch:2018aa}.
Substituting these values into Eq.~\ref{eq.doteturb} indicates that the WIM contributes only $\sim 10\%$ of the \hi\ component to $\dot{e}_{\rm turb}$.
Thus, the contributions of molecular and ionized gas to both $\dot{e}_{\rm bubble}$ and $\dot{e}_{\rm turb}$ are small enough that they do not affect the comparison between the two rates, and we therefore restrict our analysis to \hi.

The kinetic energy of the superbubbles we identify is preserved within the \hi\ disk.  
Applying M31's gas fraction ($f_{\rm gas} \sim 0.1$) and angular rotation rate ($\Omega \sim 10$ to $45\,{\rm Gyr^{-1}}$) to a general superbubble evolution model \cite{Orr:2022aa} shows that superbubbles in M31 are expected to stall and fragment within the disk.  
In our data, the superbubbles we identify have radii smaller than half the \hi\ disk thickness of $500$ to $1000\,{\rm pc}$ \cite{Braun:1991ty}. 
Their expansion velocities are comparable to or smaller than the local \hi\ velocity dispersion at their galactocentric radii (see Figs.~\ref{f.m31_profile} and~\ref{f.rvage}), a behavior also seen in nearby galaxies \cite{Bagetakos:2011ub}.  
\editfm{These signatures are consistent with the more evolved stages inferred above (and in Fig~\ref{f.mainchart} d) and indicate that the bubbles are now coupling their energy to the surrounding neutral gas.}
As a result, the bubbles remain confined to the disk and deposit their kinetic energy into the surrounding neutral gas, which makes $\dot{e}_{\rm bubble}$ a direct tracer of SN energy injection into the neutral interstellar medium. 

Our analysis resolves the local balance between SN energy injection and turbulence in M31.
While our analysis focuses on the locations where bubbles are detected, we attempt to extend these results and examine whether the superbubbles also reflect the broader pattern of SN activity in M31.
The 118 superbubbles we identify occupy a combined volume corresponding to a filling factor of $f_{\rm v} \sim 7.5$\,\% within the \hi\ disk of M31, a conservative lower limit because smaller or fainter bubbles and those embedded in complex \hi\ structures are likely missed.  
Numerical simulations of Milky Way–like galaxies predict $f_{\rm v}$ values of $40$--$60$\,\% \cite{Li:2015aa,Kim:2017ab}, implying that our observed $f_{\rm v}$ captures roughly $10$--$20$\,\% of the intrinsic porosity.  
This fraction closely matches the proportion of M31's total SN activity traced by the bubbles (10--20\,\%), which is derived from the $\sim 10^4$ SNe required to power the bubbles over their lifetimes of $10-40\,{\rm Myr}$, assuming a global SN rate of 0.2--0.4 per century in M31\cite{Caldwell:2025aa}.
In addition, the surface density of young stars enclosed within bubbles is consistent with the galaxy-wide average\cite{Kang:2009aa}, see Fig.~\ref{f.obsne} and Materials and Methods.
The bubbles we detect therefore constitute a statistically representative subset of M31’s recent SN activity despite inevitable incompleteness.
Thus, although our analysis measures the energy balance only at the locations where bubbles are identified, the same relation between $\dot{e}_{\rm bubble}$ and $\dot{e}_{\rm turb}$ should hold across the entire galaxy. 
Future facilities and improved bubble identification algorithms will further increase completeness and enable energy-balance comparisons on both local and global scales.  

Other potential drivers of turbulence do not appear to supply energy at the level required by the observed $\dot{e}_{\rm turb}$.  
The magnetorotational instability (MRI), often invoked as a contributor to galactic turbulence, is expected to be weak in M31.  
The magnetic field in the disk is $B \sim 5\,\mu\mathrm{G}$ at $R = 6$–$14\,\mathrm{kpc}$ \cite{Fletcher:2004us}, well above the MRI threshold $B_{\rm Max\,MRI} \sim 0.2$–$1\,\mu\mathrm{G}$ derived from marginal stability conditions using $\rho_{\rm HI}$ and $\Omega$ \cite{Kitchatinov:2004aa,Gressel:2013aa}.  
Using $B_{\rm Max\,MRI}$ to estimate the corresponding MRI-driven energy injection yields values 1–2 orders of magnitude below $\dot{e}_{\rm turb}$ (see Fig.~\ref{f.mainchart}a).  
We also do not find evidence in our data for large-scale \hi\ flows, another potential source of turbulent energy injection \cite{Krumholz:2017wu,Krumholz:2018aa}, although deeper or higher-resolution observations may better constrain their contribution.  

Our measurements capture the kinematic imprint of superbubbles, which provide a direct signature of SN energy injection into the interstellar medium. 
Because their expansion timescales trace the timing of feedback events and do not depend on assumptions about stellar evolution, the initial mass function, or star formation history, superbubbles offer a uniquely direct probe of recent SN activity.  
Their large spatial extent makes them resolvable in nearby galaxies, and the use of \hi\ emission enables full-disk coverage that is unaffected by dust and applicable across diverse galactic environments. 
Our results also show that superbubbles provide direct tests of stellar feedback and star-formation theory, as their momentum output and expansion timescales align with key theoretical expectations (see above).
Limitations that previously hindered superbubble studies, including insufficient angular resolution, limited sensitivity, and difficulties in robust identification, are now being overcome with new and upcoming facilities such as FAST, the Square Kilometer Array, and the Next Generation Very Large Array, together with advances in machine learning.
These developments make superbubbles a practical and scalable tracer of feedback–interstellar medium coupling, opening a path toward spatially resolved studies of feedback-regulated galaxy evolution.

\section*{Acknowledgements}
We thank M.~Krumholz and C.~Bacchini for helpful discussions, and H.~Chen for assistance with coordination.
This work used data from FAST, a Chinese national megascience facility operated by the National Astronomical Observatories of the Chinese Academy of Sciences (NAOC).
We also acknowledge the use of archival JVLA data from project 14A-235 (principal investigator A.~Leroy).
The National Radio Astronomy Observatory is a facility of the US National Science Foundation operated under cooperative agreement by Associated Universities, Inc.
Data reduction and analysis made use of the Common Astronomy Software Applications (CASA) \cite{The-CASA-Team:2022aa}, Astropy \cite{Astropy-Collaboration:2013aa,Astropy-Collaboration:2018aa,Astropy-Collaboration:2022aa} and ROHSA \cite{Marchal:2019th}.

During the subsequent revision for \textit{Nature Astronomy}, coauthor Rouyu Li independently performed a blind consistency check of the bubble classification.
This check is described in the published Version of Record (\url{https://doi.org/10.1038/s41550-026-02981-9}); it is not included in the scientific content of this initial-submission manuscript.

\section*{Funding}
This work is supported by the National Natural Science Foundation of China (NSFC), grant No.~12588202.
F.~M. is supported by NSFC grant No.~12403031.
C.-W.~T. is supported by NSFC grants Nos.~12041302 and 11988101, and CAS project No.~JZHKYPT-2021-06.
Jingwen Wu is supported by NSFC grants Nos.~12550003 and 12421003, by the National Key R\&D Program of China No.~2023YFA1608004, and the Tianchi Talent Program of Xinjiang Uygur Autonomous Region.
H.~X. is supported by NSFC grant No.~12588201.
Z.-Y.Z. acknowledges support from NSFC grants Nos.~12533003 and 1257030642.
Z.-Y.Z. is supported by the Fundamental Research Funds for the Central Universities, grant No.~KG202502.
M.-M.~M.~L. acknowledges partial support from US National Science Foundation grant AST23-07950.
D.~L. acknowledges support from the New Cornerstone Foundation and the Guizhou Leading Talent Workstation for Sci-Tech Innovation in Extreme Universe Research.

\clearpage %

\providecommand{\bibinfo}[2]{#2}

\clearpage
\renewcommand{\thefigure}{S\arabic{figure}}
\renewcommand{\thetable}{S\arabic{table}}
\renewcommand{\theequation}{S\arabic{equation}}
\renewcommand{\thepage}{S\arabic{page}}
\setcounter{figure}{0}
\setcounter{table}{0}
\setcounter{equation}{0}
\setcounter{page}{1} %

\subsection*{Materials and Methods}

\subsubsection*{Observations}
\label{sec:observations}

We combined the FAST and JVLA \hi\ data to produce a datacube sensitive to spatial scales from $\sim200$\,pc to the full extent of M31.
The JVLA D-array data, described by Ref.\ \cite{Koch:2021vf}, were obtained from the JVLA archive (project ID 14A-235; hereafter referred to as the JVLA data).
We re-imaged the visibilities with natural weighting, achieving a spatial resolution of $57$\arcsec$\times59$\arcsec.
The native spectral resolution is 0.42\,\kms, which we averaged by a factor of four during imaging to enhance the signal-to-noise ratio and reduce processing time, yielding a final spectral resolution of 1.68\,\kms.

The FAST observations were carried out between 2018 and 2022.
M31 was mapped in tracking mode for a total of 133\,h between Dec. 2018 and Aug. 2019 and again between Dec. 2021 and Jul. 2022.
An additional $\sim100$\,h of coverage was obtained in on-the-fly mode from Oct. to Dec. 2019, and a further 10\,h in drift-scan mode from Nov. to Dec. 2021.
Technical specifications of these observing modes are summarized in Ref.~\cite{Jiang:2020ua}, and a full account of the FAST observations will be presented in Zhang et al. (submitted).
The final FAST image has an angular resolution of $174$\arcsec.

\begin{figure}[!ht]  
\centering  
\includegraphics[width=0.75\textwidth]{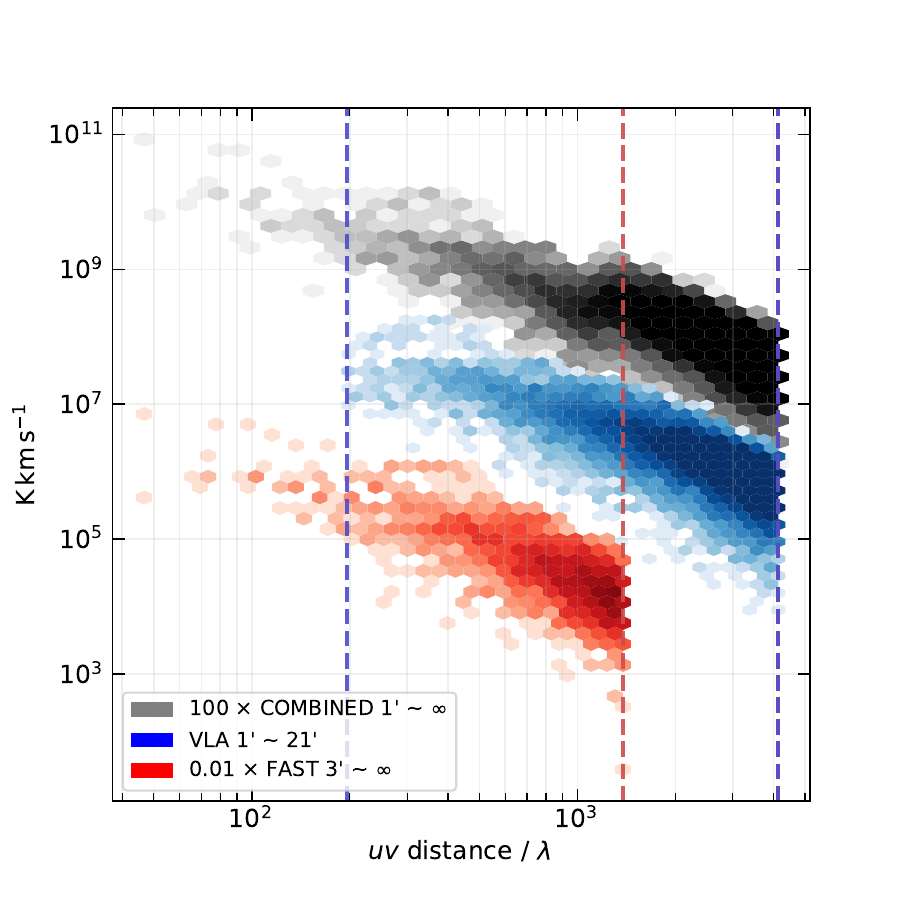}  
\caption{\textbf{FAST, JVLA, and combined image in the flux--\emph{uv} distance domain.}  
For clarity, the FAST and combined data are scaled by 0.001 and 1000, respectively.  
The blue dashed lines mark the beam size and the largest angular scale recoverable by the JVLA.  
The red dashed line marks the FAST beam.  
Sub-beam and super-JVLA angular scales are omitted from the plot.\label{f.combineuv}}  
\end{figure}

We combined the FAST and JVLA data using our J-comb method \cite{Jiao:2022up}.
This approach first matches the flux scales of the two datasets in the \emph{uv} domain.
We measured the flux ratio between the FAST and JVLA images over spatial scales from $60$\arcsec\ to $120$\arcsec\ (Fig.~\ref{f.combineuv}) and derived a correction factor of 3.26 to be applied to the JVLA visibilities.
The datasets were then combined in the image domain using J-comb, and the total flux of the resulting image was scaled to match that of the FAST map, under the assumption that FAST fully recovers emission on scales above its beam size.
The final combined datacube has a beam size of $60$\arcsec, identical to that of the JVLA image.
Adopting a distance of 744\,kpc to M31 \cite{Vilardell:2010vm}, this corresponds to a spatial resolution of 216\,pc.
The final RMS of the combined cube is $\sim2$\,mJy\,beam$^{-1}$ per 1.68\,\kms\ channel, corresponding to a column-density sensitivity of $N_{\rm HI} \approx 2.65\times10^{17}$\,cm$^{-2}$.
Figure~\ref{f.combineuv} illustrates the flux–\emph{uv}-distance distributions for the FAST, JVLA, and combined datasets.

\subsubsection*{Bubble Identification}
Visual identification remains one of the most robust approaches for detecting \hi\ bubbles in complex environments, despite the availability of automated methods \cite{Daigle:2007tb,Ehlerova:2005ul,Ehlerova:2013uh,Watkins:2023wj}.
In this study, we adopted visual identification supplemented by a filtering procedure to remove false detections.
FM inspected the cube channel by channel following the morphological criteria of Ref.~\cite{Brinks:1986vn}, selecting candidates solely on the basis of morphology and visual contrast.
Only structures exhibiting partially elliptical features that persist across at least three consecutive channels (corresponding to $\sim5$\,\kms) were retained, yielding an initial list of $\sim200$ candidates.

\begin{figure*}[!ht]  
\centering  
\includegraphics[width=0.95\textwidth]{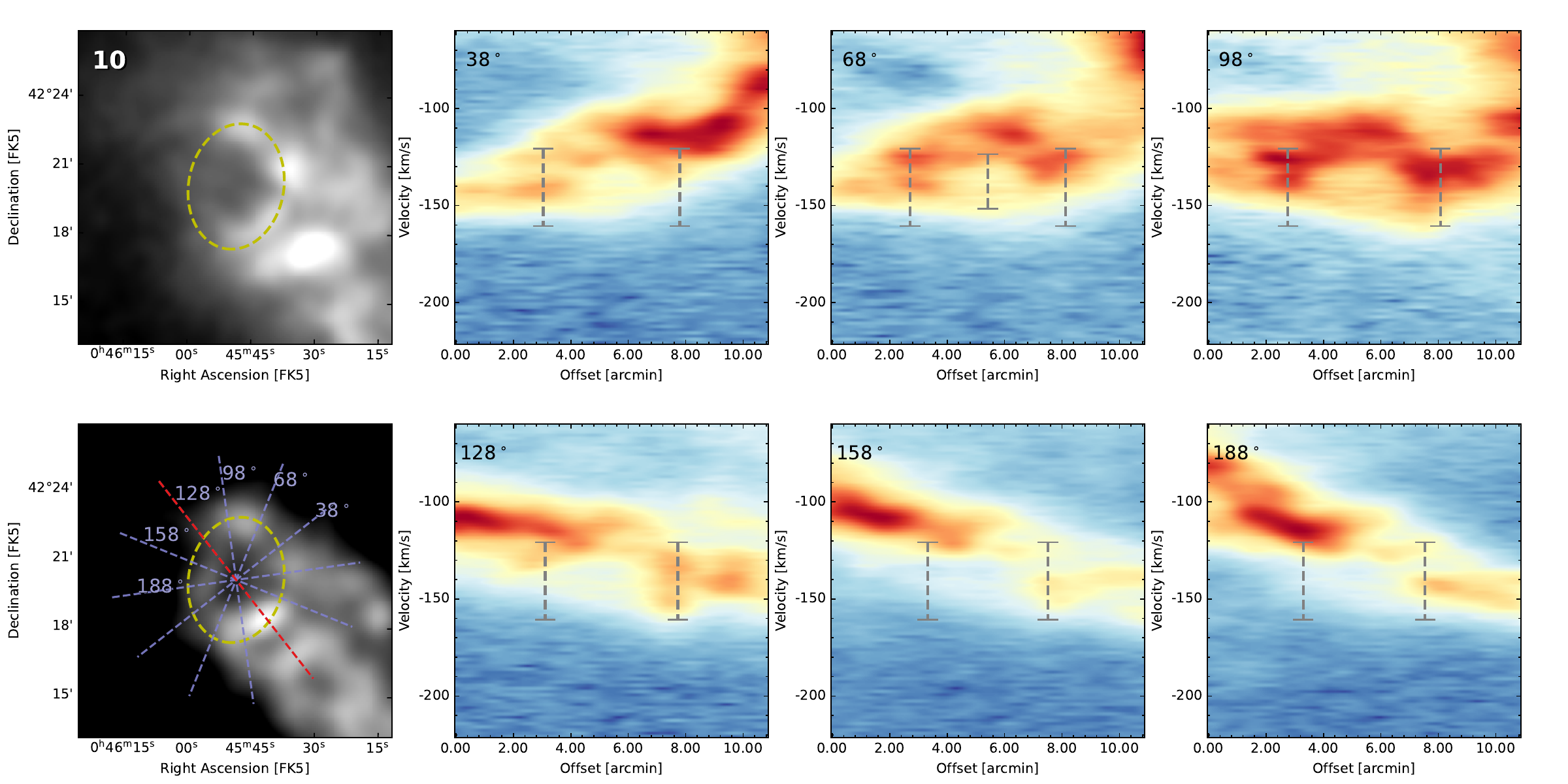}\\  
\includegraphics[width=0.95\textwidth]{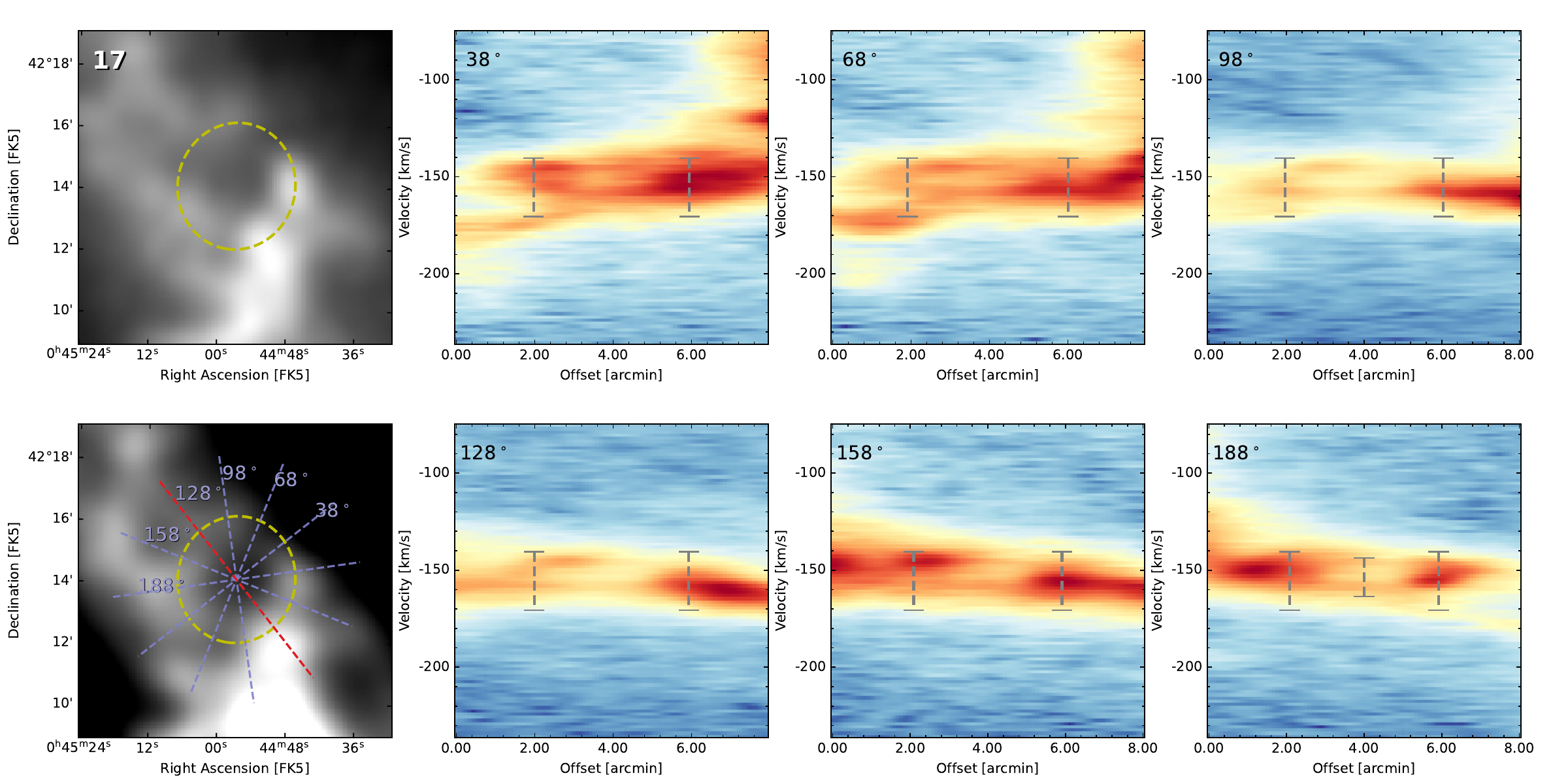}  
\caption{\textbf{Two examples of bubble identification (\#10 and \#17).}\label{f.type_i_bubbles}  
For each bubble, the upper left panel shows the channel map at the velocity $(v_{\rm max}+v_{\rm min})/2$, while 
the lower left panel shows the moment 0 map over the velocity range $(v_{\rm min},\,v_{\rm max})$.  
The bubble's shape ($\theta_{\rm maj},\, \theta_{\rm  min},\,{\rm PA}$) is indicated by yellow dashed ellipses.  
The blue dashed segments show the PV cut directions.  
The short dashed lines in the 68$^\circ$ PV plot of bubble \#27 and in the 188$^\circ$ plot of bubble \#17 mark the $\delta v$ measurements.  
The longer dashed segments in each PV plot indicate the velocity range $(v_{\rm min},\,v_{\rm max})$ used for mass estimates and moment 0 extraction.  
Their horizontal positions correspond to the projected location of the bubble edge in each direction.}  
\end{figure*}

We then generated position--velocity (PV) diagrams for each candidate along six directions.
Two of these directions were aligned parallel and perpendicular to the major axis of M31, and the other four were evenly spaced in position angle (PA).
These PV diagrams allow the direct identification of expanding structures, which manifest as elliptical or partial ring-like signatures in PV space.
An ideal, isotropically expanding bubble appears as a complete ellipse, although the axis ratio may differ due to the differing spatial and velocity scales.
We kept only candidates that show clear and continuous full or partial elliptical signatures in at least one of the six PV diagrams.
For incomplete ellipses, we further required that the expansion velocity be measurable along the visible arc.
Applying these criteria, we obtained a final sample of 118 bubbles with well-defined expansion signatures.
Figure~\ref{f.type_i_bubbles} presents two representative examples of bubble identification in PV space.

The combined use of position--position morphology and PV-based kinematic signatures effectively suppresses false positives.
The dominant source of false negatives is the data quality.
Given our spatial resolution of $\sim200$\,pc, bubbles with \editfm{$r\lesssim200$\,pc} are likely unresolved and therefore missed.
Highly elongated or asymmetric structures may also evade morphological selection.
In addition, expanding components below the detection threshold, for example those with column densities $N_{\rm HI}<3\times{\rm RMS}\approx8\times10^{17}$\,cm$^{-2}$ per 1.68\,\kms\ channel, would not be identified.
We therefore regard the 118 bubbles in our sample as a conservative lower bound on the number of detectable expanding structures in our data.

\subsubsection*{Observational Properties of the Bubbles}
\label{sec:bubble_properties}

We characterized each of the 118 superbubbles in M31 by measuring their basic structural and kinematic properties.
For every bubble, we recorded the right ascension and declination of its center, the semi-major axis $\theta_{\rm maj}$, semi-minor axis $\theta_{\rm min}$, the PA of the best-fit ellipse, and the velocity range over which the bubble is detected ($v_{\rm min}$ to $v_{\rm max}$).

We determined these quantities through a three-step measurement procedure.
First, we identified approximate central coordinates from a channel map where the bubble is clearly visible.
Second, we generated position--velocity (PV) diagrams along six directions and selected the one with the clearest expansion signature to define $v_{\rm min}$ and $v_{\rm max}$.
This velocity range was adjusted to include all \hi\ components plausibly associated with the bubble, and its midpoint corresponds to the channel where the bubble is most distinct.
Because the velocity interval merely brackets the bubble’s appearance, the true expansion velocity $v_{\rm exp}$ is always smaller than $(v_{\rm max} - v_{\rm min})/2$.
Third, using either the moment-0 map integrated over $(v_{\rm min},\,v_{\rm max})$ or the central channel map, whichever better reveals the morphology, we refined the bubble center and measured $\theta_{\rm maj}$, $\theta_{\rm min}$, and PA through elliptical fitting.

\begin{figure*}[!ht]
\centering
\includegraphics[width=0.95\textwidth]{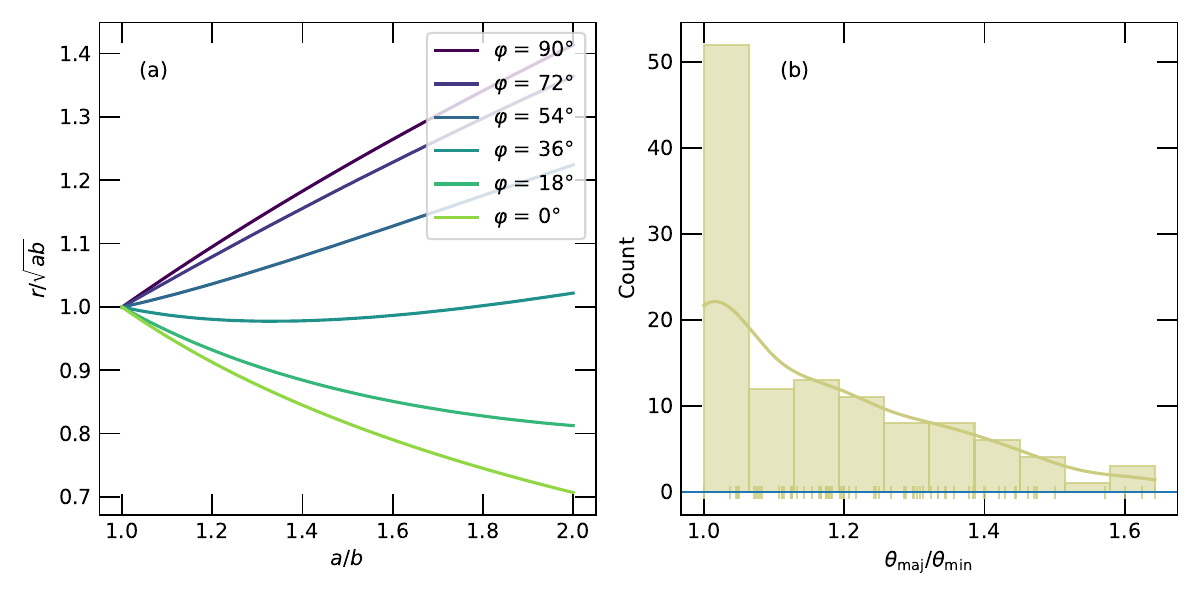}
\caption{\textbf{Accuracy of using $r$ to represent $\sqrt{ab}$ and the distribution of axis ratios.}  
(a) Predicted values of $r/\sqrt{ab}$ for ellipses with varying intrinsic axis ratios $a/b$ and orientation angles $\varphi$  
between the ellipse major axis $\vec{a}$ and the major axis of M31.  
Here, $r$ denotes the projected semi-major axis.  
(b) Distribution of observed axis ratios $\theta_{\rm maj}/\theta_{\rm min}$ in our sample.  
Since the intrinsic values of $a/b$ are unknown, this distribution is used as a proxy to estimate the likely range of $a/b$.  
\label{f.aoverb}}
\end{figure*}

We estimated the physical bubble radius $r$ from the observed semi-major axis $\theta_{\rm maj}$.
For a spherical bubble, the projected major axis directly corresponds to the true radius regardless of inclination.
For intrinsically elliptical structures with semi-major axes $a$ and $b$, the projected $\theta_{\rm maj}$ depends on the PA $\varphi$ and axis ratio $a/b$.
To test the robustness of using $\theta_{\rm maj}$ as a proxy for $\sqrt{ab}$, we evaluated the ratio $r/\sqrt{ab}$ across a range of $a/b$ and $\varphi$.
As shown analytically in Fig.~\ref{f.aoverb}a, this ratio remains between 0.8 and 1.2 for $a/b<1.5$ and is nearly symmetric around unity.
Because $\sim95\%$ of our bubbles satisfy $\theta_{\rm maj}/\theta_{\rm min}<1.5$ (Fig.~\ref{f.aoverb}b), we adopt a spherical approximation and assign a 20\% geometric uncertainty to $r$.

We measured the expansion velocity $v_{\rm exp}$ from the separation of the approaching and receding limbs in the selected PV diagram.
This definition provides a good approximation for isotropic expansion.
Given the smooth edges in PV space and the subjective nature of this measurement, we assigned a typical uncertainty of 20\%.

We computed the galactocentric radius $R$ of each bubble by deprojecting its position relative to the center of M31.
Using an inclination of $i=77^\circ$ and major-axis PA $\phi_0=38^\circ$ \cite{Corbelli:2010wx}, and letting $(\alpha_0,\delta_0)$ denote the galaxy center while $(\alpha,\delta)$ denotes a given bubble, we calculated
\begin{equation}
    R = R_* \left[ \cos\phi + \frac{\sin\phi}{\cos i} \right]^{1/2},
\end{equation}
where
\begin{equation}
    R_* = \frac{216\,{\rm pc}}{{\rm arcmin}} \left[ (\alpha - \alpha_0)^2 + (\delta - \delta_0)^2 \right]^{1/2},
\end{equation}
and
\begin{equation}
    \phi = \arctan\left( \frac{\alpha - \alpha_0}{\delta - \delta_0} \right) - \phi_0.
\end{equation}

\begin{figure*}[!ht]
\centering
\includegraphics[width=0.75\textwidth]{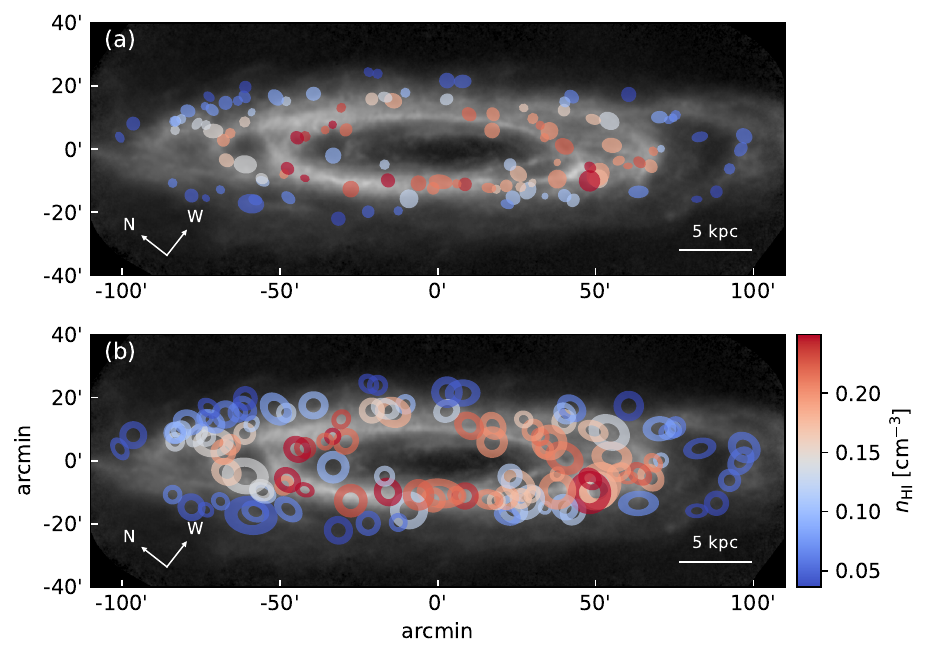}
\caption{
\textbf{How we measured $n_{\rm HI}$.}\label{f.nhi_ring_preview}
Panel (a) shows the rotated \hi\ moment zero map with the superbubble ellipses overlaid, corresponding to the regions used to define the interior of each bubble.  
Panel (b) shows the same background with elliptical ring regions overlaid, constructed as annuli between the original bubble ellipses and ellipses whose semi-major axes are scaled by a factor of two.  
In both panels the ellipses and rings are colored by the ring averaged \hi\ number density $n_{\rm HI,ring}$ (cm$^{-3}$).  
}
\end{figure*}

We propagated uncertainties in the galactocentric radius $R$ by accounting for both disk thickness and bubble size.
The total uncertainty is
\begin{equation}
    \delta R = \sqrt{(h \tan i \sin\phi)^2 + r^2},
\end{equation}
where $h$ is the local scale height.
For example, a bubble at $R=15\,{\rm kpc}$ with $r=400\,{\rm pc}$ has $\delta R \sim 0.4$–$1.8\,{\rm kpc}$ depending on viewing angle.

We estimated the ambient hydrogen density $n_{\rm HI}$ for each bubble from the ring-averaged \hi\ column density around it and the local scale height of the disk (see Fig.~\ref{f.nhi_ring_preview}).
Specifically, we constructed an elliptical annulus by scaling the bubble ellipse from radius $r$ to $2r$ and masking out both the bubble interior itself and all other identified bubbles before computing the mean $N_{\rm HI}$ within the annulus.
This ring geometry is designed to sample the surrounding \hi\ that represents the pre-expansion ambient medium while excluding emission from the evacuated cavity and from neighboring bubbles.
We then converted the ring-averaged column density to a characteristic volume density using $n_{\rm HI} = 2\,N_{\rm HI}/h$, where $h$ is the local scale height from \cite{Braun:1991ty} and the factor of 2 assumes an approximately symmetric vertical distribution.
This estimate carries a typical uncertainty of $\sim20\%$, dominated by uncertainties in $h$ and departures from vertical homogeneity.

We estimated the total uncertainty in the bubble energy injection rate $\dot{e}_{\rm bubble}$ by propagating fractional uncertainties of $\sim20\%$ in $r$, $v_{\rm exp}$, and $n_{\rm HI}$.

\subsubsection*{Galactic Shear}

Differential rotation in galactic disks can mimic expansion signatures in PV space, producing false bubble detections \cite{Wada:2002aa}.
To avoid this contamination, we quantified the maximum apparent velocity splitting that can be produced purely by galactic shear at the location of each bubble.
Following the Oort $A$ formalism and assuming a flat rotation curve with $V_0 = 220\,{\rm km\,s^{-1}}$ \cite{Zhang:2024wn}, we computed the projected shear limit
\begin{equation}
    v_{\rm lim,proj} \simeq (V_0/R)\,r\,\sin i\,|\cos\alpha|,
\end{equation}
where $i$ is the disk inclination, and $\alpha$ the in-plane azimuthal angle ($\alpha=0$ or $\pi$ along the major axis and $\alpha=\pi/2$ or $3\pi/2$ along the minor axis).

\begin{figure*}[!ht]
\centering
\includegraphics[width=0.55\textwidth]{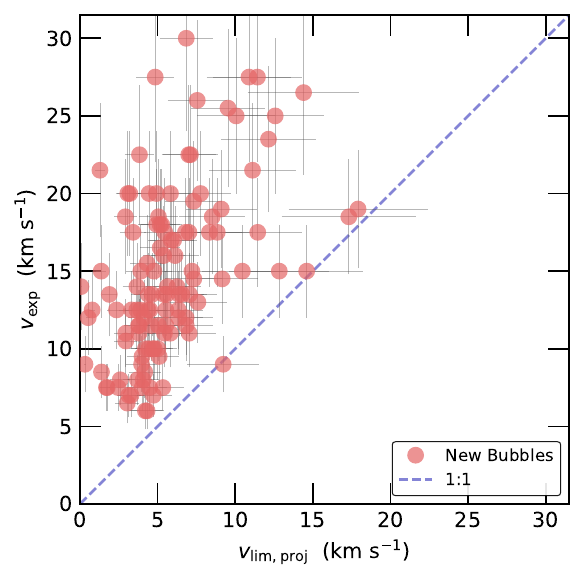}
\caption{
\textbf{Observed bubble expansion velocities compared with the projected shear limit in M31.}
Each bubble is plotted with its observed expansion velocity $v_{\rm exp}$ against the projected shear limit $v_{\rm lim,proj}$ (blue dashed line).
The projected shear limit is evaluated as  $v_{\rm lim,proj} = (V_0 / R)\, r\, \sin i\, |\cos\alpha|,$
assuming a flat rotation curve with $V_0 = 220\,{\rm km\,s^{-1}}$ \cite{Zhang:2024wn}.
Here $i$ is the disk inclination and $\alpha$ the in-plane azimuth ($\alpha=0$ or $\pi$ along the major axis and $\alpha=\pi/2$ or $3\pi/2$ along the minor axis).
\label{f.nsf}
}
\end{figure*}

We compared the observed expansion velocities to this shear limit to determine whether the measured $v_{\rm exp}$ can be accounted for by differential rotation alone.
Figure~\ref{f.nsf} shows this comparison, where the dashed line marks $v_{\rm exp} = v_{\rm lim,proj}$ and points are colored by the estimated number of SNe per bubble.

All bubbles in our sample have $v_{\rm exp}$ well above $v_{\rm lim,proj}$, indicating that the observed velocity splittings cannot be attributed to galactic shear.
This confirms that the expansion signatures are genuine and require SN-driven bubble growth rather than rotational kinematics.

\subsubsection*{Supernova Count and Momentum Injection per Supernova}

\begin{figure*}[!ht]
\centering{
\textsf{
\includegraphics[width=0.55\textwidth]{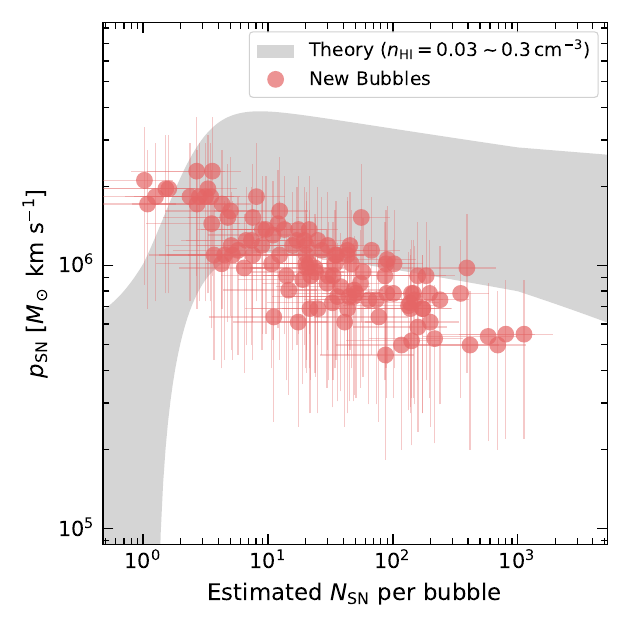}
\caption{\textbf{Momentum per SN $p_{\rm SN}$ versus the estimated number of SNe per bubble $N_{\rm SN}$.}  
Circles mark bubbles evaluated with the Weaver model \cite{Weaver:1977wh}%
.  
The gray shaded region shows a theoretical prediction of $p_{\rm SN}$ \cite{Gentry:2017aa}, calculated for ambient hydrogen densities $n_{\rm HI} = 0.05$--$0.4\,{\rm cm^{-3}}$.  
\label{f.momentum}}  
}}
\end{figure*}

\begin{figure*}[!ht]
\centering{
\includegraphics[width=0.55\textwidth]{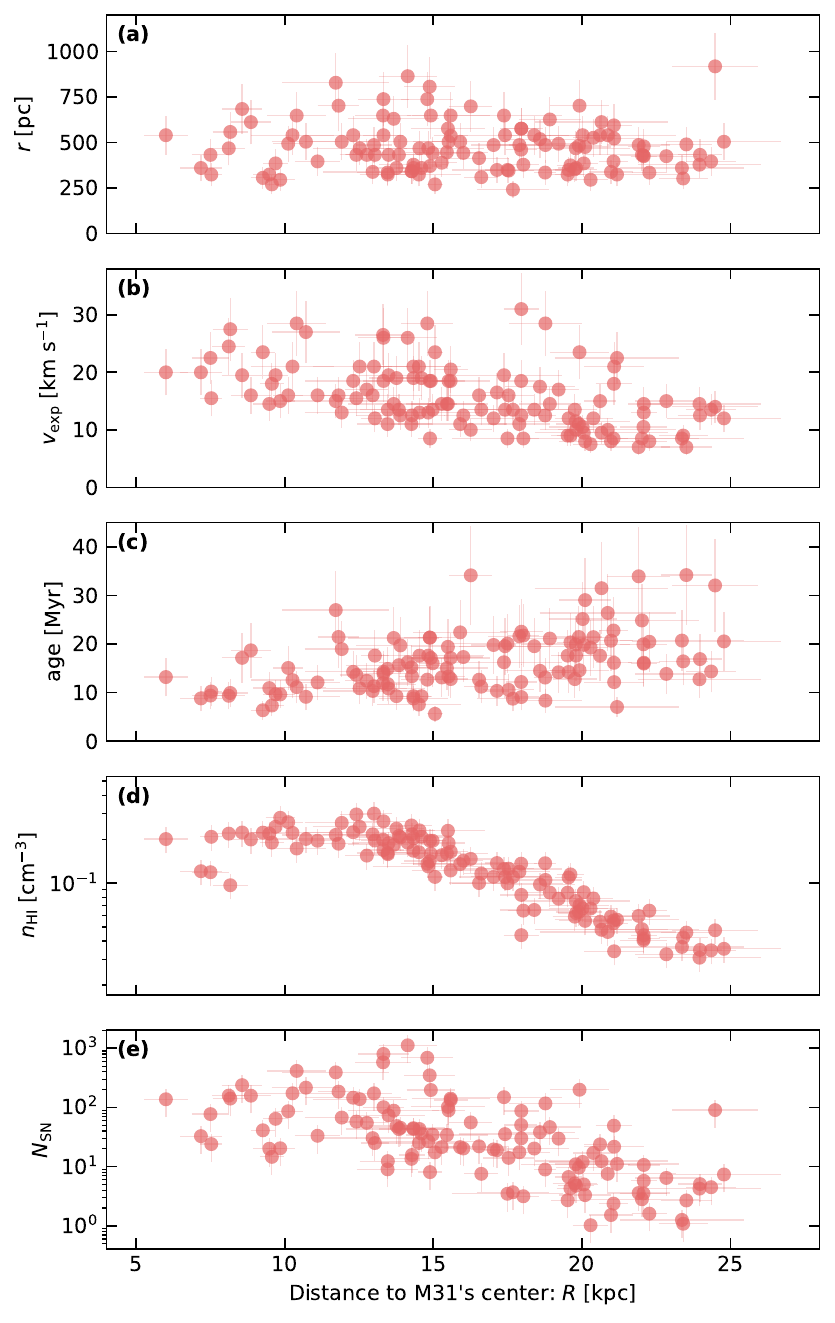}
\caption{
\textbf{Superbubble properties as a function of galactocentric radius $R$ in M31.\label{f.rvage}}
(a) Radius $r$.
(b) Expansion velocity $v_{\rm exp}$.
(c) Kinematic age $t_{\rm exp} = 0.5\,r/v_{\rm exp}$.
(d) Ambient \hi\ density $n_{\rm HI}$.
(e) Number of supernovae $N_{\rm SN}$ required to drive each bubble \cite{Weaver:1977wh}.
}
}
\end{figure*}

To estimate the number of SNe required to drive each superbubble, we adopt the Weaver model \cite{Weaver:1977wh}.
For a bubble expanding into a uniform medium with ambient hydrogen density $n_{\rm HI}$, radius $r$, and expansion velocity $v_{\rm exp}$, the Weaver solution implies a total mechanical energy of the form
\begin{equation}
E_{\rm Weaver} \approx 6\times10^{50}\,\mathrm{erg}\,
\left( \frac{v_{\rm exp}}{10~\mathrm{km\,s^{-1}}} \right)^{2}
\left( \frac{r}{100~\mathrm{pc}} \right)^{3}
\left( \frac{n_{\rm HI}}{1~\mathrm{cm^{-3}}} \right).
\end{equation}
The Weaver model assumes that the bubble interior retains nearly all injected mechanical energy.
Simulations that include thermal conduction and turbulent mixing \cite{El-Badry:2019tf} show that a significant fraction of the injected energy is radiated away at the interface between the hot interior and the swept-up shell.
For ambient densities comparable to those of our bubbles, with a typical value of $n_{\rm HI}\approx0.1~\mathrm{cm^{-3}}$, these simulations find that only a fraction $(1-\theta)\approx0.4$--$0.6$ of the input energy remains available to drive the expansion, corresponding to an overall correction factor of approximately $1/(1-\theta)\approx 2$.
We therefore define the required injected energy as
\begin{equation}
E_{\rm total}
= \frac{E_{\rm Weaver}}{1-\theta}
\;\approx\;
2\,E_{\rm Weaver}.
\end{equation}
The number of SNe required to power each bubble then follows from
\begin{equation}
N_{\rm SN} = \frac{E_{\rm total}}{E_{\rm SN}},
\end{equation}
where we adopt $E_{\rm SN}=10^{51}\,\mathrm{erg}$ as the fiducial mechanical energy per SN.
Using this procedure, the superbubbles in our sample require total injected energies spanning $10^{51}$--$10^{53}\,\mathrm{erg}$, corresponding to a few to a few hundred SNe (see Fig.~\ref{f.rvage}), demonstrating that they are produced by multiple sequential SNe rather than single explosions.

For each bubble we also compute the momentum contributed per SN by dividing its measured momentum by the number of SNe derived above.
Using $p_{\rm bubble} = (4\pi/3)\,r^{3}\,n_{\rm HI}\,\mu\,v_{\rm exp}$, the momentum per SN is
\begin{equation}
p_{\rm SN}
= \frac{4\pi}{3}\,\frac{r^{3}\,n_{\rm HI}\,\mu\,v_{\rm exp}}{N_{\rm SN}},
\end{equation}
with all quantities defined in the main text.
The resulting values span $p_{\rm SN}\approx10^{5.5\text{--}6.5}\,M_{\odot}\,{\rm km\,s^{-1}}$, consistent with theoretical expectations for SN-driven bubbles evolving in gas with densities comparable to those in the M31 \hi\ disk \cite{Kim:2015aa,Kim:2017ab,Gentry:2017aa}.
The full distribution of $p_{\rm SN}$ versus $N_{\rm SN}$ is shown in Fig.~\ref{f.momentum}, where the gray shaded band indicates the range predicted by one-dimensional numerical simulations \cite{Gentry:2017aa}.

\subsubsection*{OB-Star Sampling and Comparison with Bubble-Based Supernova Estimates}

\begin{figure*}[!ht]
\centering{
\includegraphics[width=0.55\textwidth]{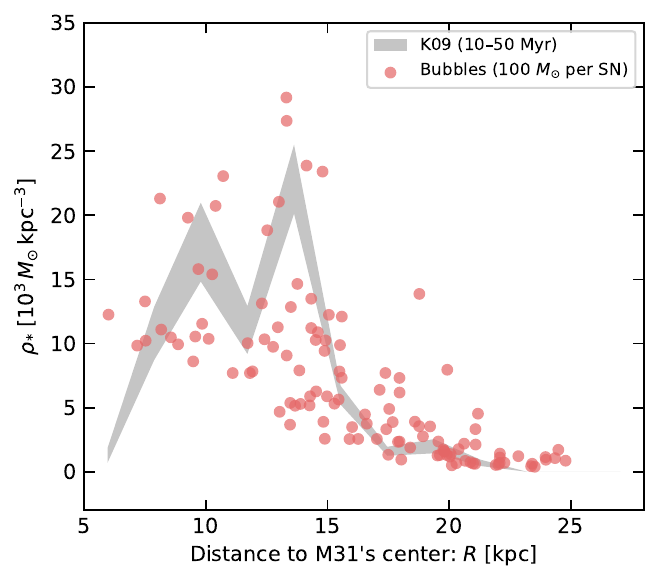}
\caption{
\textbf{Comparison between the stellar mass densities of star-forming regions and superbubbles in M31.}  
The grey curve shows the stellar mass density estimated from 284 star-forming regions in the OB-star census of M31 \cite{Kang:2009aa} with ages between 10 and 50\,Myr, whereas the red circles indicate the stellar mass density inferred for individual \hi\ superbubbles from the number of SNe required to power them, assuming $100\,M_{\odot}$ of stars formed per SN \cite{Zapartas:2017aa}. 
The shading represents the Poisson uncertainty of the binned stellar populations.
\label{f.obsne}.
}}
\end{figure*}

To test whether the SN rates implied by the superbubbles are consistent with the underlying young stellar population in M31, we compared our results with the catalog of star-forming regions identified from resolved OB stars \cite{Kang:2009aa}.
From this catalog we selected regions with ages between $10$ and $50$\,Myr, matching the typical expansion timescales of our superbubbles.

For a consistent comparison with the bubble-based estimates, we converted stellar mass to an expected number of SNe using $100\,M_\odot$ per SN for a Kroupa initial mass function \cite{Hopkins:2006aa,Zapartas:2017aa}.
We then normalized the OB-star-derived stellar masses by the volume of the \hi\ disk at the corresponding galactocentric radii, yielding volumetric SN rate estimates directly comparable to those inferred from the superbubbles (Fig.~\ref{f.obsne}).

The radial profiles derived from the OB-star census and from the superbubbles show similar absolute levels and parallel declines with radius.
This agreement demonstrates that the SN activity traced by the bubbles reflects the underlying young stellar population across M31, rather than a biased subset driven by selection effects.

\subsubsection*{Turbulent Energy Estimation}
\label{sec:turbulent_energy_estimation}

We estimate the kinetic energy density of \hi\ turbulence using
\[
e_{\rm turb} = \tfrac{3}{2}\,\mu_{\rm H}\,n_{\rm HI}\,\sigma_{\rm turb}^2,
\]
where $\mu_{\rm H} n_{\rm HI}$ is the local mass density and $\sigma_{\rm turb}$ is the one-dimensional turbulent velocity dispersion.

For each pixel we compute the observed line width from the second moment of the \hi\ spectrum.
We first identify the longest contiguous velocity interval in which the emission exceeds $2\times$ the RMS noise measured in line-free channels, pad this interval by three channels on both sides, and mask all other channels.
The observed dispersion is
\[
\sigma_{\rm obs}
= \sqrt{\frac{\sum I(v)v^{2}}{\sum I(v)}
       - \left(\frac{\sum I(v)v}{\sum I(v)}\right)^{2}},
\]
where $I(v)$ is the masked line intensity.
To isolate the turbulent component, we correct for thermal broadening in the warm neutral medium (WNM) using a fractional WNM contribution of $f_{\rm WNM}\approx0.6$ and a WNM thermal dispersion of $\sigma_{\rm th}\approx 8\,\mathrm{km\,s^{-1}}$ \cite{Dickey:1988ue}:
\[
\sigma_{\rm turb}
= \sqrt{\sigma_{\rm obs}^{2} - f_{\rm WNM}\,\sigma_{\rm th}^{2}}.
\]

\begin{figure*}[!ht]
\centering{
\includegraphics[width=0.55\textwidth]{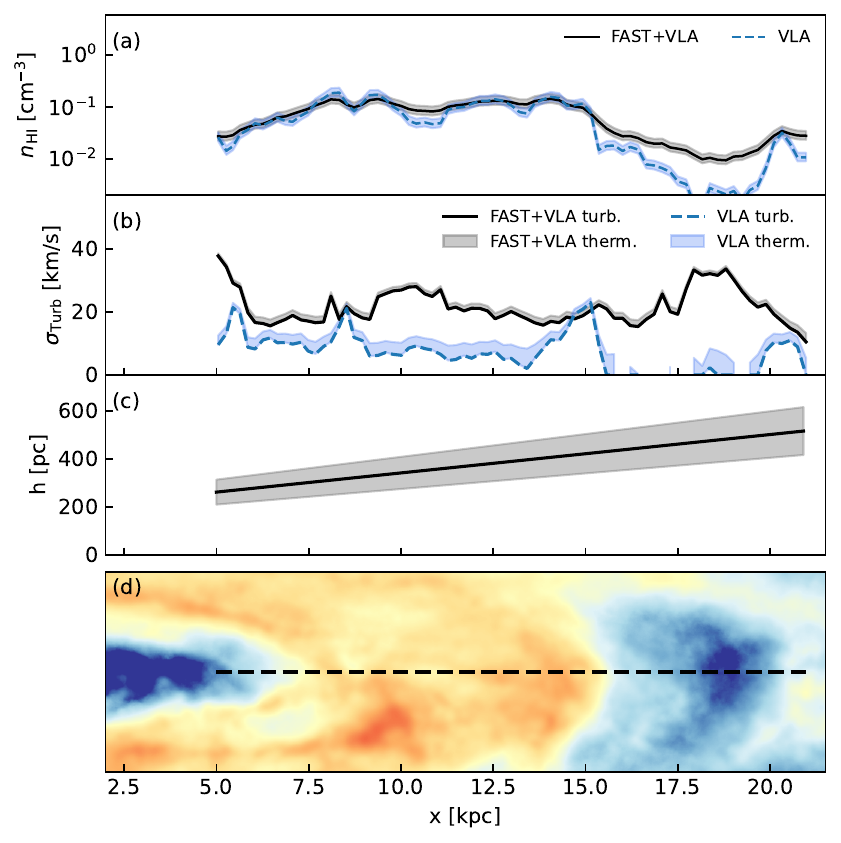}
\caption{
\textbf{Profiles of physical parameters along the southern major axis of M31\label{f.m31_profile}.}  
(a) Number density profile corrected for projection effects.  
The FAST+JVLA dataset used in our analysis is shown as a black solid curve, and the JVLA-only dataset as a blue dashed curve; shaded regions indicate uncertainties.  
(b) Velocity dispersion ($\sigma_{v}$) profile, with thermal broadening indicated by shading.  
Black solid and blue dashed curves again represent the FAST+JVLA and JVLA-only datasets, respectively.  
(c) Scale height $h$ of the H\,\textsc{i} disk \cite{Braun:1991ty}; 
(d) Moment-0 map with the profile extraction path marked.  
}}
\end{figure*}

The inclusion of the FAST data is essential for reliably estimating $\sigma_{\rm turb}$ and hence $\dot{e}_{\rm turb}$, as it recovers diffuse emission with intrinsically broad profiles that the JVLA alone resolves out \cite{Yue:2021aa,Plunkett:2023aa}.
The combined FAST+JVLA dataset, which retains the JVLA-only angular resolution, exhibits \hi\ velocity dispersions that are approximately twice those derived from the JVLA data alone (see Fig.~\ref{f.m31_profile}), ensuring that the measured line widths reflect the true turbulent motions of the gas.

\begin{figure*}[!ht]
\centering
\includegraphics[width=0.98\textwidth]{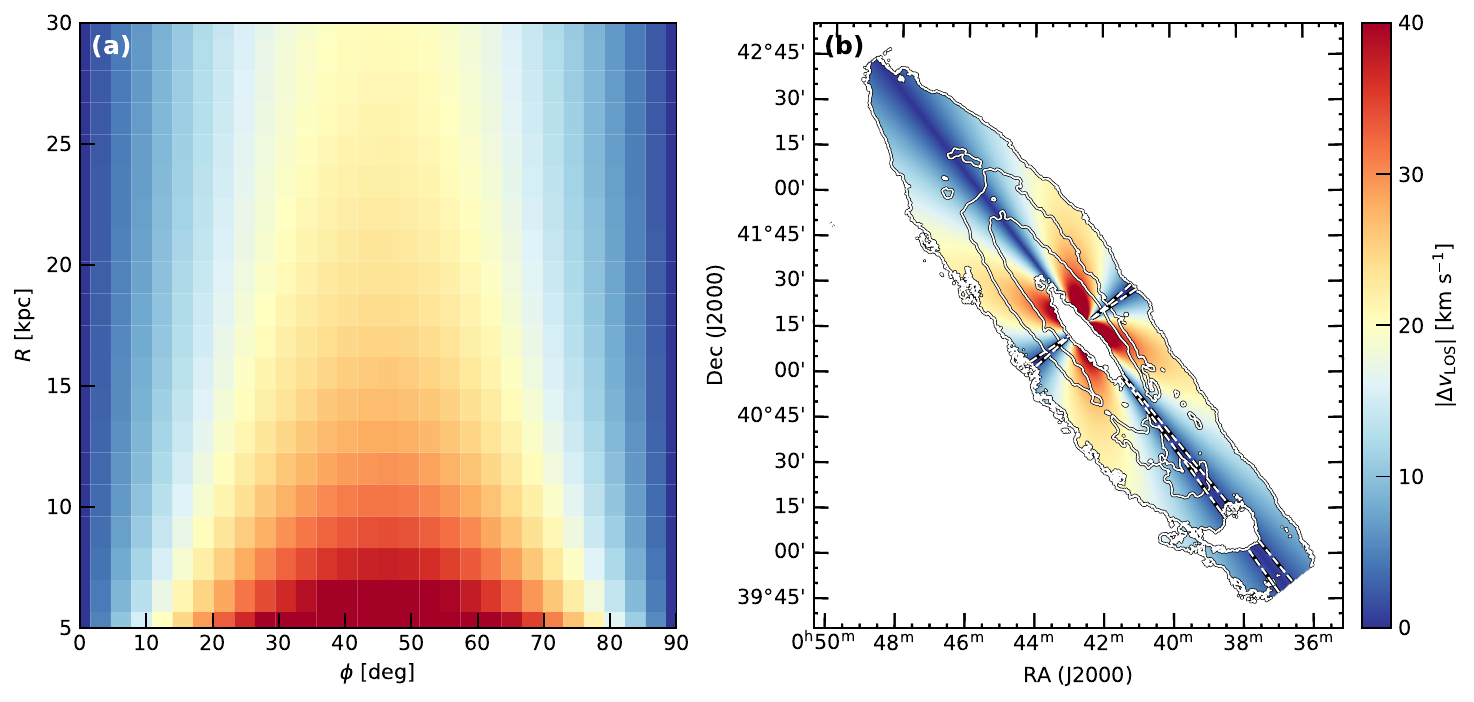}
\caption{
\textbf{Geometric model of projection-induced velocity broadening and its manifestation on the sky.}
(\textbf{a})~Model prediction for the projection-induced velocity broadening $\Delta v_{\rm LOS}$ as a function of azimuthal angle $\phi$ and galactocentric radius $R$.
We assume an inclined \hi\ disk ($i = 77^\circ$) with vertical thickness \cite{Braun:1991ty} $h/({1\, \rm pc}) = 182 + 16\,R/(1\, {\rm kpc})$,
and compute the geometric velocity spread caused by the projection effect of a finite thick disk along each line of sight.
The broadening is largest at intermediate $\phi$ and vanishes along both the major ($\phi = 0^\circ,\,180^\circ$) and minor ($\phi = 90^\circ,\,270^\circ$) axes.
(\textbf{b})~Map of $|\Delta v_{\rm LOS}|$, same as (a) but in projected RA and Dec frame, 
overlaid with \hi\ column density contours at $N_{\rm HI} = 5\times10^{20}$ and $2\times10^{21},{\rm cm^{-2}}$.
Dashed white segments mark where $\Delta v_{\rm LOS} = 1.67\,{\rm km\,s^{-1}}$, corresponding to the spectral resolution limit.
Three narrow azimuthal sectors with $\Delta v_{\rm LOS} < 1.67\,{\rm km\,s^{-1}}$ were used to calculate the turbulence dissipation rate $\dot{e}_{\rm turb}$.
The northern major-axis sector was excluded to avoid contamination from foreground Milky Way emission.
}
\label{f.deltav_model}
\end{figure*}

Because M31 is highly inclined ($i = 77^\circ$), a single line of sight intersects a finite vertical extent of the disk.
Gas at different height $z$ samples different in-plane azimuthal angles $\phi$ and therefore different projected rotation velocities.
This geometric effect produces an intrinsic LOS velocity broadening that is distinct from classical beam smearing and depends only on the disk thickness and viewing geometry.

We quantify this projection-induced broadening by computing the angular range $\delta\phi$ subtended by the disk thickness $h$ at galactocentric radius $R$:
\[
\delta \phi = 
\arctan\!\left(\frac{R \sin\phi + \tan i \cdot h}{R \cos\phi}\right)
-
\arctan\!\left(\frac{R \sin\phi - \tan i \cdot h}{R \cos\phi}\right),
\]
with an associated LOS velocity spread
\[
\delta v \approx v_{\rm rot}\,\sin i\,\sin\phi\,\delta\phi.
\]
This formulation captures the geometrical projection effects and, to our knowledge, has not been previously quantified for \hi\ in disk galaxies.

For estimating $\dot{e}_{\rm turb}$, we retain only those pixels for which the modeled projection-induced broadening is smaller than the spectral resolution ($\delta v < 1.7\,\mathrm{km\,s^{-1}}$), ensuring that the measured $\sigma_{\rm turb}$ reflects intrinsic turbulence.
The selected pixels naturally form four elongated strips around the major and minor axes, where projection effects vanish or minimize.
The northern major-axis strip is excluded to avoid contamination from Galactic foreground emission.

Figure~\ref{f.m31_profile} shows the resulting radial profiles along the southern major axis: (a) $n_{\rm HI}$, (b) $\sigma_{\rm turb}$ (with the thermal component shaded), (c) the adopted scale height $h$ \cite{Braun:1991ty}, and (d) the moment-0 map with the extraction path.
Typical uncertainties are $\sim 20\%$ in both $n_{\rm HI}$ and $\sigma_{\rm turb}$ and $\sim 20$--$30\%$ in $h$, giving a propagated uncertainty of $\sim 50\%$ in $\dot{e}_{\rm turb}$.

\subsubsection*{Molecular-to-Atomic Gas Ratios Around Superbubbles}

\begin{figure*}[!ht]
\centering{
\includegraphics[width=0.55\textwidth]{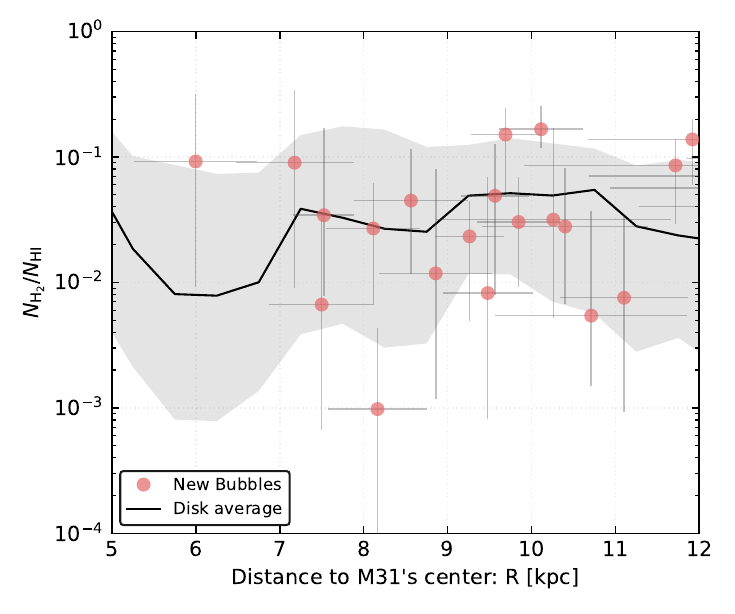}
\caption{
\textbf{Radial profile of the molecular-to-atomic gas ratio in M31 and local values around superbubbles.}
The black line shows the azimuthally averaged $N_{\rm H_2}/N_{\rm HI}$ ratio with the gray band indicating the 25th–75th percentile range in 0.5 kpc radial bins.
The $N_{\rm H_2}$ map is derived from CO(1--0) data using a conversion factor $X_{\rm CO}=1.9\times10^{20}\ {\rm cm^{-2}\,(K\,km\,s^{-1})^{-1}}$ \cite{Nieten:2006tr}.
Red circles show individual superbubbles, where the local molecular fraction is measured as the median $N_{\rm H_2}/N_{\rm HI}$ within an elliptical ring surrounding each bubble and plotted at the bubble center radius.
Horizontal error bars reflect geometric uncertainties in $R$, and vertical error bars span the 25th–75th percentile range within each ring.
\label{f.moltohiratio}
}
}
\end{figure*}

To characterize the ambient conditions in which the superbubbles reside, we measured the local molecular-to-atomic gas ratio $N_{\rm H_2}/N_{\rm HI}$ around each bubble and compared these values to the azimuthally averaged radial profile in M31 (Fig.~\ref{f.moltohiratio}).
The $N_{\rm H_2}$ map is derived from the CO(1--0) data of \cite{Nieten:2006tr} using a conversion factor $X_{\rm CO}=1.9\times10^{20}\ {\rm cm^{-2}\,(K\,km\,s^{-1})^{-1}}$, and the atomic column densities come from the combined FAST+JVLA \hi\ data.

For each superbubble, we compute the local molecular fraction as the median $N_{\rm H_2}/N_{\rm HI}$ within an elliptical ring surrounding the bubble, chosen to sample the ambient medium while excluding the interior cavity.
The radial position of each point corresponds to the bubble center, and horizontal error bars reflect the geometric uncertainty in galactocentric radius from disk thickness and bubble size.
Vertical error bars represent the interquartile range of the molecular fraction within the ring.
Circle colors encode the ambient atomic gas density $n_{\rm HI}$.

\end{document}